\documentstyle[12pt]{article}
\title{\bf
Real-time thermal field theory analyses of 2D Gross-Neveu model\thanks{
This work was partially supported by National Natural Science Foundation of
China and by Grant No. LWTZ-1298 of Chinese Academy of Sciences.}}
\author{
{\bf 
ZHOU Bangrong } \\
\normalsize Department of Physics, Graduate School at Beijing \\
\normalsize University of Science and Technology of China,
  Academia Sinica, Beijing 100039, {\bf China} \\
\normalsize and  \\
\normalsize CCAST (World Laboratory), Beijing 100080, {\bf China} \\
}
\date {}
\begin{document}
\hoffset = -1 truecm
\voffset = -2 truecm
\baselineskip = 12pt
\maketitle

\begin{abstract}
Discrete symmetry breaking and possible restoration at finite temperature $T$
are analysed in 2D Gross-Neveu model by the real-time thermal field theory in
the fermion bubble approximation. The dynamical fermion mass $m$ is proven to be
scale-independent and this fact indicates the equivalence between the fermion
bubble diagram approximation and the mean field approximation used in the
auxialiary scalar field approach. Reproducing of the non-zero critical
temperature $T_c=0.567 m(0)$, ($m(0)$ is the dynamical fermion mass at $T=0$),
shows the equivalence between the real-time and the imaginary-time thermal
field theory in this problem.  However, in the real-time formalism, more results
including absence of scalar bound state, the equation of criticality curve of
chemical potential-temperature and the $\ln(T_c/T)$ behavior of $m^2$ at
$T\stackrel{<}{\sim} T_c$ can be easily obtained.  The last one indicates the
second-order phase transition feature of the symmetry restoration.
\end {abstract}
PACS numbers: 11.10.Kk, 11.10.Wx, 11.30.Er, 11.30.Qc \\
Key words: 2D Gross-Neveu model,
           spontaneous discrete symmetry breaking,
           real-time thermal field theory,
           fermion bubble diagram approximation
%%-main body of paper-%%

\section{Introduction}
In 4D model of NJL-form [1] and 3D Gross-Neveu (GN) model [2], by means of
analyses of dynamical fermion mass in the real-time thermal field theory in
the fermion bubble approximation, we have acquired a quite clear understanding
of restorations of the spontaneously broken symmetries at high temperature and
their second-order phase transition feature [3,4].  A natural idea is to extend
the same analysis to 2D GN model. However, once entering one-space dimension,
one will find some special features. By the Mermin-Wagner-Coleman Theorem [5],
a continuous symmetry can never be spontaneously broken in 2 dimensions.
Consequently we will
have to confine ourselves to possible spontaneous breaking of only some
discrete symmetries. The 2D GN model is perturbatively renormalizable and
asymptotically free and this is apparently different from the perturbatively
non-renormalizable 4D NJL model and 3D GN model. Based on the auxialiary scalar
field approach, many researches on the behavior of 2D GN model at finite
temperature have been made [6,7,8], though all of them in the imaginary-time
formalism of thermal field theory.  Considering the above facts, in this paper,
we will analysed 2D GN model in the real-time thermal field theory in the
fermion bubble approximation.  Through comparison between the results obtained
by our approach and the known ones, we will be able to check the equivalence
of the real-time and the imaginary-time formalism of thermal field theory in
this problem, find the advantages of the real-time formalism and the limitations
of the fermion bubble approximation and , in the meantime, get deeper insight
into the 2D GN model itself as well.\\
\indent The paper is arranged as follows. In Sect.2 we will analyse the
symmetries of 2D GN model, mainly its discrete symmetries and their possible
spontaneous breaking induced by dynamical fermion mass. In Sect. 3 the
behavior of the model at temperature $T=0$ will be outlined. This includes
derivation of the renormalized gap equation and the running coupling constant
in dimension regularization scheme and the proof of that scalar bound state
could not exist in this model. In Sect. 4 we extend the gap equation to the case
of $T\neq 0$, determine the critical temperature of symmetry restoration and
derive the critical behavior of the dynamical fermion mass.  Finally in Sect. 5
we indicate the advantages and the limitations of our approach through the
comparison with the known results and reach corresponding conclusions.
\section{Symmetry analysis of the model}
The Lagrangian of 2D GN model is expressed by
$${\cal L}=\sum_{k=1}^N\bar{\psi}^ki\gamma^{\mu}\partial_{\mu}
\psi_k+\frac{g}{2}\sum_{k=1}^N[\bar{\psi}^k\psi_k]^2.
\eqno(2.1)$$
\noindent where $\psi\equiv \psi_k(t,x)$ are the spinor fields with $N$ "color"
components. In 2 dimesions, the four-fermion coupling constant $g$ is
dimensionless thus the theory is perturbatively renormalizable. The
$\gamma^{\mu} (\mu=0,1)$ and $\gamma_5=\gamma^0 \gamma^1$ are $2\times 2$
matrices which are related to the Pauli matrices $\sigma^i (i=1,2,3)$ by
$$\gamma^0=\sigma^1=\left(\matrix{0 & 1 \cr
                                  1 & 0 \cr}
                    \right),
  \gamma^1=i\sigma^2=\left(\matrix{0 & 1 \cr
                                  -1 & 0 \cr}
                    \right)  \ \ {\rm and}
\gamma_5=-\sigma^3=\left(\matrix{-1 & 0 \cr
                                  0 & 1 \cr}
                    \right)
\eqno(2.2)$$
\noindent i.e. we take the "Weyl" representation.  In this representation,
$\psi_k$ will be two-component complex spinors.  We will concern only the
discrete symmetries of the model, since continuous symmetries can not be
spontaneously broken in 2 dimensions.  It is easy to see that the Lagrangian
(2.1) is invariant under the discrete transformation $R$,
$$\psi (t,x)\stackrel{R}{\longrightarrow} -\psi (t,x)
\eqno(2.3)$$
\noindent and the ordinary parity ${\cal P}$ and the time reversal ${\cal T}$,
$$\psi (t,x)\stackrel{{\cal P}}{\longrightarrow} \gamma^0\psi (t,-x),$$ $$
  \psi (t,x)\stackrel{{\cal T}}{\longrightarrow} \gamma^0\psi (-t,x).
\eqno(2.4)$$
\noindent In addition, it is also invariant under the discrete chiral
transformation
$\chi_D$,
$$\psi (t,x)\stackrel{\chi_D}{\longrightarrow} \gamma_5\psi (t,x)
\eqno(2.5)$$
and the special parity ${\cal P}_1$,
$$\psi (t,x)\stackrel{{\cal P}_1}{\longrightarrow} \gamma^1\psi (t,-x)
\eqno(2.6)$$
\noindent We indicate that the fermion mass term $-m\bar{\psi}\psi$ will keep
$R$, ${\cal P}$ and ${\cal T}$ invariant but not $\chi_D$ and ${\cal P}_1$
invariant.
Therefore, the dynamical generation of the fermion mass will imply spontaneous
breaking of the discrete chiral symmetry $\chi_D$ and the special parity
${\cal P}_1$. The spontaneous breaking of $\chi_D$ and ${\cal P}_1$ and their
possible restoration at finite temperature $T$ is just the problem we will
discussed.
\section{The behavior of the model at $T=0$}
The renormalization operation will be conducted in the dimension regularization
approach [9]. For this purpose, we change the time-space dimension from 2 into
$$D=2-2\varepsilon
\eqno(3.1)$$
\noindent and the Lagrangian ${\cal L}$ in Eq.(2.1) into
$${\cal L}_D=\sum_{k=1}^N\bar{\psi}^ki\gamma^{\mu}\partial_{\mu}
\psi_k+\frac{g}{2}M^{2-D}\sum_{k=1}^N[\bar{\psi}^k\psi_k]^2,
\eqno(3.2)$$
\noindent where $M$ is a scale parameter with mass dimension, the mass
dimensions $[g]=0$, $[\psi_k]=[\bar{\psi}^k]=(D-1)/2$ and the dimension of
the representation of the matrices $\gamma^{\mu}$ becomes $2^{D/2}$. The
Feynman rule of the four-fermion vertex will be $igM^{2-D}$. \\
\indent Assuming that the four-fermion interactions will lead to the
condensates $\sum_{k=1}^N\langle\bar{\psi}^k\psi_k\rangle\neq 0$, then the
non-zero dynamical mass
$$m(0)=-gM^{2-D}\sum_{k=1}^N\langle\bar{\psi}^k\psi_k\rangle
\eqno(3.3)$$
\noindent will be generated and the discrete symmetries $\chi_D$ and
${\cal P}_1$ will be spontaneously broken. After the Wick rotation and
introduction of the denotations $\bar{l}^0=il^0, \bar{l}^i=l^i$, we obtain
from Eq.(3.3) the gap equation
\begin{eqnarray*}
1&=&gN2^{D/2}M^{2-D}\int_{-\infty}^{\infty}
   \frac{d^D\bar{l}}{(2\pi)^D}\frac{1}{\bar{l}^2+m^2(0)} \\
 &=&gN2^{D/2}M^{2-D}\frac{\Gamma(1-D/2)}{(4\pi)^{D/2}[m^2(0)]^{1-D/2}},
\end{eqnarray*}
$$\eqno(3.4)$$
\noindent where we have used the integral formula [10]
$$\int_{-\infty}^{\infty}
   \frac{d^D\bar{l}}{(2\pi)^D}\frac{1}{[\bar{l}^2+m^2(0)]^A}
 =\frac{\Gamma(A-D/2)}{(4\pi)^{D/2}\Gamma(A)[m^2(0)]^{A-D/2}}.
\eqno(3.5)$$
\noindent Considering Eq. (3.1), the gap equation can be reduced to
$$1=\frac{gN}{2\pi}\left[
     \frac{1}{\varepsilon}-\gamma+\ln\frac{2\pi M^2}{m^2(0)}+
     {\cal O}(\varepsilon)\right]
\eqno(3.6)$$
\noindent where $\gamma =0.5772$ is the Euler constant. The divergent
$1/\varepsilon$ term can be removed by introduction of the counter-term of
the four-fermion interactions
$${\cal L}_{c.t.}=\frac{g}{2}M^{2-D}(Z_4-1)\sum_{k=1}^N
                  (\bar{\psi}^k\psi_k)^2.
\eqno(3.7)$$
\noindent The renormalization constant $Z_4$ is defined by the minimal
subtraction scheme [9], i.e., to one-loop order,
$$Z_4=1-\frac{gN}{2\pi}\frac{1}{\varepsilon},
\eqno(3.8)$$
\noindent then the renormalized gap equation will take the form
$$1=\frac{gN}{2\pi}\left[
  \ln\frac{2\pi M^2}{m^2(0)}-\gamma \right]
\eqno(3.9)$$
\noindent Eq. (3.9) gives that
$$m^2(0)=2\pi M^2\exp[-\frac{2\pi}{gN}-\gamma]
\eqno(3.10)$$
\noindent Since 2D GN model is an asymptotically free theory, as will be argued
immediately, the $g$ in Eq. (3.10) should be replaced by corresponding running
coupling constant.  The running coupling can be derived by the general
renormalization group method.  In fact, after introducing the counter-term
${\cal L}_{c.t.}$, we can define the bare four-fermion coupling constant
$$g_0=gM^{2-D}Z_4
\eqno(3.11)$$
\noindent which should be independent of the renormalization scale parameter
M, i.e.
$$\frac{dg_0}{dM}=0
\eqno(3.12)$$
\noindent Considering Eq. (3.8), Eq. (3.12) will lead to
$$M\frac{\partial g}{\partial M}=-\frac{N}{\pi}g^2
\eqno(3.13)$$
\noindent and furthermore,
$$\frac{1}{g(M)}=\frac{N}{\pi}\ln \frac{M}{M_0}+\frac{1}{g(M_0)}
\eqno(3.14)$$
\noindent Eq. (3.14) indicates that the running coupling $g(M)\to 0$ when
$M\to \infty$, i.e. the theory is ultraviolet stable and asymptotically free.
Now we replace the coupling constant $g$ in Eq. (3.10) by the running coupling
$g(M)$, then by means of Eq. (3.14), can prove that
$$m^2(0)=2\pi M^2 \exp[-\frac{2\pi}{g(M)N}-\gamma]
        =2\pi M_0^2 \exp[-\frac{2\pi}{g(M_0)N}-\gamma]
\eqno(3.15)$$
\noindent i.e. the dynamical fermion mass $m(0)$ is scale-indepedent, we will
obtain the same $m(0)$ in the scales $M$ and $M_0$.  On the other words, the
order parameter $m(0)$ of the discrete symmetry breaking will be independent
of the size of the system. \\
\indent We can also calculate the renormalized fermionic four-point function
$\Gamma_S^R(p)$ in scalar channel by the methods used in Ref. [11]. The result
is
$$\Gamma_S^R(p)=-i/[p^2-4m^2(0)+i\varepsilon]K_0(p),
\eqno(3.16)$$
$$K_0(p)=\frac{N}{4\pi}\int_{0}^{1}\frac{dx}{m^2(0)-p^2x(1-x)}
        =\frac{N}{\pi p^2 \sqrt{4m^2(0)/p^2-1}}
                \arctan \frac{1}{\sqrt{4m^2(0)/p^2-1}},
\eqno(3.17)$$
\noindent where we have made the assumption that $0<p^2<4m^2(0)$. To examine if
$p^2=4m^2(0)$ could be a simple pole of $\Gamma_S^R(p)$, we may calculate the
residual of $\Gamma_S^R(p)$ at $p^2=4m^2(0)$ and obtain
$$\lim \limits_{p^2\to 4m^2(0)}\left[
   \frac{4m^2(0)}{p^2}-1\right]\Gamma^R_S(p)=0
\eqno(3.18)$$
\noindent which implies that $\Gamma^R_S(p)$ does not represent a propogator
for any scalar bound state with the mass $2m(0)$. Hence, the scalar bound state
does not exist at $T=0$ in this model.  In fact, this conclusion can be
generalized to the case of $T\neq 0$, as will be showed in the next Section.
\section{Symmetry restoration and phase transition at finite temperature}
We will take the real-time formalism of thermal field theory [12] to research
the temperature behavior of the dynamical fermion mass.  This means that,
corresponding to the vacuum expectation value
$\sum_{k}\langle\bar{\psi}^k\psi_k\rangle$ replaced by the thermal expectation
value $\sum_{k}{\langle\bar{\psi}^k\psi_k\rangle}_T$, we should make the
substitution of the fermion propagator
$$\frac{i}{\not\!{l}-m+i\varepsilon}\longrightarrow
  \frac{i}{\not\!{l}-m+i\varepsilon}-2\pi\delta(l^2-m^2)
  (\not\!{l}+m)\sin^2\theta(l^0,\mu),$$ $$
\sin^2\theta(l^0, \mu)=\frac{\theta(l^0)}{\exp[\beta(l^0-\mu)]+1}
                        +\frac{\theta(-l^0)}{\exp[\beta(-l^0+\mu)]+1},
\eqno(4.1)$$
\noindent where $\mu$ is chemical potential and $\beta=1/T$.  Let $m\equiv
m(T,\mu)$ is the dynamical fermion mass at finite temperature $T$ and finite
chemical potential $\mu$, then similar to the derivation at $T=0$, we may obtain
the gap equation at $T\neq 0$
$$1=Z_4\frac{gN}{2\pi}\left[
     \frac{1}{\varepsilon}-\gamma+\ln\frac{2\pi M^2}{m^2}
     -2\int d^2l \ \delta(l^2-m^2)\sin^2\theta(l^0,\mu)
     \right]
\eqno(4.2)$$
\noindent Notice that the integral related to $T$ is convergent and the UV
divergence appears only in the zero temperature sector. Therefore, by the
definition (3.8) of $Z_4$ at $T=0$, we can remove the $1/\varepsilon$
divergence to the one-loop order and obtain the renormalized gap equation
at $T\neq 0$
$$1=\frac{gN}{2\pi}\left\{
     \ln\frac{2\pi M^2}{m^2}-\gamma
     -2[I_1(y,-r)+I_1(y,r)] \right\}
\eqno(4.3)$$
\noindent where
$$I_1(y,\mp r)=\int_{0}^{\infty}\frac{dx}{\sqrt{x^2+y^2}}
             \frac{1}{\exp(\sqrt{x^2+y^2}\mp r)+1}
\eqno(4.4)$$
\noindent with the denotations $y=m/T$ and $r=\mu/T$. By the same argument as
the one leading to Eq. (3.15), it can be proven that the dynamical fermion
mass $m$ at finite temperature and chemical potential is also
scale-independent. This implies that in the bubble fermion diagram
approximation, the order parameter is always size-independent. \\
\indent Substitute the gap equation (3.9) at $T=0$ into Eq. (4.3), we will
obtain
$$\ln \frac{m(0)}{m}=I_1(y,-r)+I_1(y,r)
\eqno(4.5)$$
\noindent We may assume that there is a critical temperature $T_c\neq 0$ so
that when $T\to T_c$, $m\to 0$. It is easy to see that this assumption is
consistent with Eq. (4.5).  In fact, when $m\to 0$, in the left-handed side
of Eq. (4.5) there will appear $\ln \infty$ divergence and in the
right-handed side, the integral at $x\to 0$ will approach $\int_{0}^{\infty}
\frac{dx}{x}$ and has the same $\ln \infty$ behavior.  Thus the assumption
$T_c\neq 0$ is plausible, at least in the bubble diagram approximation.\\
\indent Since when $T\stackrel{<}{\sim} T_c\neq 0$, $m\sim 0$, i.e. $y\ll 1$,
we can use the high temperature expansion of $I_1(y, \mp r)$ [3] and obtain
from Eq. (4.5)
$$\ln\frac{m(0)}{T\pi}+\gamma=\frac{7}{2}\zeta (3)(\frac{y}{2\pi})^2
  +\left[7\zeta (3)-93\zeta (5)(\frac{y}{2\pi})^2 \right](\frac{r}{2\pi})^2
  $$ $$
  +\left[-31 \zeta (5)+\frac{1905}{2}\zeta (7)(\frac{y}{2\pi})^2 \right]
   (\frac{r}{2\pi})^4 +127\zeta (7)(\frac{r}{2\pi})^6+
   {\cal O}\left[(\frac{y}{2\pi})^4\right]
\eqno(4.6)$$
\noindent where $\zeta (3)=1.202,\ \zeta (5)=1.037, \ ...$ and $r/2\pi<1$
is also assumed. We find that the $\ln m$ terms in both sides of Eq. (4.6)
have cancelled each other. Taking $y=0$ i.e. $m=0$, we will obtain the
equation of $T_c$ and the critical chemical potential $\mu_c$
$$\ln\frac{m(0)}{T_c\pi}+\gamma=7\zeta (3)(\frac{r_c}{2\pi})^2
 -31\zeta (5)(\frac{r_c}{2\pi})^4 +127\zeta (7)(\frac{r_c}{2\pi})^6+\cdots
\eqno(4.7)$$
\noindent with the denotation $r_c=\mu_c/T_c$.  When chemical potential is
zero, Eq. (4.7) gives
$$T_c=(e^{\gamma}/\pi)m(0)=0.567 \ m(0)
\eqno(4.8)$$
\noindent This result is identical to the one obtained in the mean field
approximation of the auxialiary scalar field approach in the imaginary-time
formalism of thermal field theory [6-8]. When  the chemical
potential is not equal to zero, Eq. (4.7) will be able to determine a
$\mu_c-T_c$ criticality curve at $r_c/2\pi <1$. This shows an advantage of the real-time
thermal field theory approach.  Besides this, we may further give the critical
behavior of the dynamical fermion mass $m$ i.e. the order parameter of symmetry
breaking at $T\stackrel{<}{\sim} T_c$. Since $y\ll 1$ when
$T\stackrel{<}{\sim} T_c$, it is reasonable to keep only the terms of the order
${\cal O}(y^2)$ in Eq. (4.6). Then, considering the equation (4.7) of
$\mu_c-T_c$ curve (in which $\mu_c$ is redenoted by $\mu$), we may derive the
expression for the squared dynamical fermion mass
$$m^2=4\pi^2T^2\left[\ln\frac{T_c}{T}+
      7\zeta (3){\left(\frac{\mu}{2\pi}\right)}^2\left(
      \frac{1}{T_c^2}-\frac{1}{T^2}\right)-
      31\zeta (5){\left(\frac{\mu}{2\pi}\right)}^4\left(
      \frac{1}{T_c^4}-\frac{1}{T^4}\right)\right.$$ $$
      \left.+127\zeta (7){\left(\frac{\mu}{2\pi}\right)}^6\left(
      \frac{1}{T_c^6}-\frac{1}{T^6}\right)  \right]/
      \left[\frac{7}{2}\zeta (3)-93\zeta (5){\left(\frac{\mu}{2\pi}\right)}^2
           \frac{1}{T^2}+
           \frac{1905}{2}\zeta (7){\left(\frac{\mu}{2\pi}\right)}^4
           \frac{1}{T^4}\right],$$
           $$\ \ \ \ \ \ \ \ \ \ \ \ \ \ \ \ \ \ \ \ \ \ \ \ \ \  \ \
           {\rm when} \ \ T\stackrel{<}{\sim} T_c.
 \eqno(4.9)$$
\noindent If the chemical potential is zero, then Eq. (4.9) gives
$$m^2=\frac{8\pi^2}{7\zeta (3)}T^2\ln\frac{T_c}{T}, \ \ {\rm when} \ \
                                        T\stackrel{<}{\sim} T_c.
\eqno(4.10)$$
\noindent The logrithmic behavior of $m^2$ for $T_c$ can be compared with
the temperature behaviors of $m^2$ at $T\stackrel{<}{\sim} T_c$ in 4D NJL model
and 3D GN model. The latters are [3,4]
$$m^2=\frac{\pi^2}{3}\frac{T_c^2-T^2}{\ln(\Lambda/T\pi)+\gamma-1/2},
\ \ {\rm for \ 4D \ NJL \ model}
\eqno(4.11)$$
$$m^2=8(\ln 2)T(T_c-T), \ \ {\rm for \ 3D \ GN \ model}
\eqno(4.12)$$
\noindent In both cases we have $m\sim (T_c-T)^{1/2}$ at
$T\stackrel{<}{\sim} T_c$ and it means that the relevant symmetry restorations
are the second-order phase transitions. In the case of 2D GN model, based on
a similar argument to the one given in Ref. [3], we can reach the same
conclusion, i.e. the discrete symmetry restoration at $T>T_c$ is also the
second-order phase transition. \\
\indent Based the method used in Ref. [13], we can obtain the renormalized
fermionic four-point function $\Gamma_{ST}^R(p)$ in scalar channel at finite
temperature $T$ which is expressed by
$$\Gamma_{ST}^R(p)=-i/(p^2-4m^2+i\varepsilon)\left\{
              K(p)+H(p)-iS(p)-R^2(p)/[K(p)+H(p)+iS(p)]\right\},
\eqno(4.13)$$
\noindent where $K(p)$ has the same form as $K_0(p)$ in Eq. (3.17), except that
$m(0)$ must be replaced by m, and
$$H(p)=\frac{N}{2 \pi}\int d^2l \ \left[
         \frac{p^2-2l\cdot p}{(p^2-2l\cdot p)^2+\varepsilon^2}+
         \frac{p^2+2l\cdot p}{(p^2+2l\cdot p)^2+\varepsilon^2}\right]
         \delta(l^2-m^2)\sin^2\theta(l^0, \mu),
\eqno(4.14)$$
$$S(p)=\frac{N}{2}\int d^2l \ \delta(l^2-m^2)\delta[(l+p)^2-m^2]
       [\sin^2\theta(l^0,\mu)\cos^2\theta(l^0+p^0,\mu)+
        \cos^2\theta(l^0,\mu)\sin^2\theta(l^0+p^0,\mu)],
\eqno(4.15)$$
$$R(p)=\frac{N}{4}\int d^2l \ \delta(l^2-m^2)\delta[(l+p)^2-m^2]
       \sin2\theta(l^0,\mu)\sin2\theta(l^0+p^0,\mu).
\eqno(4.16)$$
\noindent The form (4.13) of $\Gamma_{ST}^R(p)$ ensures mutual cancellation
of the pinch singularities contained in $S(p)$ and $R(p)$. It is easy to
confirm that when $p^2\to m^2$, $H(p)$ is finite but $S(p)$ and $R(p)$ are
sigular.  The latter comes from the feature of the product of the two
$\delta$-functions. Since $K(p)$ is divergent when $p^2 \to m^2$, we will
still have the residual of $\Gamma_{ST}^R(p)$ at $p^2=4m^2$
$$\lim \limits_{p^2\to 4m^2}\left(
   \frac{4m^2}{p^2}-1\right)\Gamma_{ST}^R(p)=0
\eqno(4.17)$$
\noindent Hence, the conclusion that there is no scalar bound state in the
model will be maintained at finite temperature. \\
\indent Our results about finite temperature behavior of the model are obtained
by the real-time thermal field theory in the fermion bubble diagram
approximation.  It should be pointed that the fermion bubble diagram
approximation has already been able to reflect physical essentiality in 4D and
3D case. However, it has great limitations in 2D case.  As illustrated in
Ref. [8], where the auxialiary scalar field approach was used, for an infinite
one-space system and a finite $N$, in fact there is not symmetry restoration at
$T\neq 0$ owing to the special kink effects when the minimal configuration
$\sigma_c(x)$ of the effective thermodynamic potential
$\Gamma(\sigma_c(x),\beta)$ could be a function of space. Only if $\sigma_c(x)=$
constant is assumed, i.e. in so called the mean field approximation, one could
obtain symmetry restoration at $T\neq 0$ and this is just the result obtained
in Refs. [6,7] and in this paper.  It may be indicated that the
scale-independece, or the size-independence of the order parameter $m(T,\mu)$
proved in this paper is just consistent with the assumption $\sigma_c(x)=$
constant in the mean field approximation. Hence we can say that the fermion
bubble approximation is indeed the mean field approximation named in Ref. [8].
Since the mean field approximation is valid only if $N\to \infty$ or if $N$ is
large and one consider only a finite segment of one-space dimension system,
the results about finite temperature obtained in this paper are applicable only
in the above conditions.
\section{Conclusions}
We have analysed spontaneous breaking and restoration of the discrete chiral
symmetry $\chi_D$ and the special parity ${\cal P}_1$ induced by dynamical
fermion mass in renormalizable and asymptotically free 2D Gross-Neveu model
by means of the real-time thermal field theory in the fermion bubble diagram
approximation.  We have proven that the dynamical fermion mass, both $m(0)$
at zero-temperature and $m$ at finite temperature, are scale-independent,
consistent with the fundamental assumption of the mean field approximation
in the auxialiary scalar field approach. Hence, the fermion bubble diagram
approximation is equivalent to the mean field approximation and must submit
to the same applicable conditions. All the results, especially reproducing
of the non-zero symmetry restoration temperature $T_c=0.567 \ m(0)$, show the
equivalence between the real-time and the imaginay-time formalism of thermal
field theory in this problem.  However, in the real-time formalism, one can
further prove that there is no scalar bound state in the model, easily obtain
the equation of criticality curve of chemical potential and temperature and the
$\ln(T_c/T)$ behavior of the dynamical squared fermion mass  $m^2$ at
$T\stackrel{<}{\sim} T_c$. The last one clearly indicates the second-order
phase transition feature of the symmetry restorations.

\end{document}